\documentclass[10pt,twocolumn,a4paper]{article}
 
\usepackage[utf8]{inputenc}
\usepackage[T1]{fontenc}
\usepackage[english]{babel}
\IfFileExists{lmodern.sty}%
  {\usepackage{lmodern}\usepackage{microtype}}%
  {\usepackage[expansion=false]{microtype}}

\usepackage[hyphens]{url}

\usepackage[a4paper,margin=1.9cm]{geometry}
\usepackage{amsmath,amssymb,amsthm}
 
\usepackage{graphicx}
\graphicspath{{figures/}}
\usepackage{booktabs}
\usepackage{tabularx}
\newcolumntype{Y}{>{\centering\arraybackslash}X} 
\usepackage{multirow}
\usepackage{float}
\usepackage[table]{xcolor}
\usepackage{pifont}
\usepackage{caption} 
\usepackage{enumitem}
 
\definecolor{linkblue}{RGB}{0,90,200}
 
\usepackage[breaklinks]{hyperref}
\hypersetup{
  colorlinks=true,
  linkcolor=linkblue,   
  citecolor=linkblue,   
  urlcolor=linkblue,    
  filecolor=linkblue,
  linktoc=all,
  pdfborder={0 0 0}
}
 
\usepackage{orcidlink}
 
\usepackage{url}
\usepackage{cleveref}
\crefname{equation}{Eq.}{Eqs.}
\Crefname{equation}{Eq.}{Eqs.}

\newcommand{\doilink}[1]{%
  \href{https://doi.org/#1}{doi:\ \nolinkurl{#1}}%
}

\definecolor{editpurple}{RGB}{122,0,182}

\definecolor{q21red}{RGB}{190,0,0}
\newcommand{\eqcite}[1]{#1} 
 
\newcommand{\yes}{\ding{51}}   
\newcommand{\notmark}{\ding{55}} 
\newcommand{\nomark}{\textendash}
 
\newcommand\blfootnote[1]{%
  \begingroup
  \renewcommand\thefootnote{}\footnote{#1}%
  \addtocounter{footnote}{-1}%
  \endgroup
}
 
\newcommand{\msk}{\mathit{sk}}
\newcommand{\aid}{\mathit{id}}
\newcommand{\ppid}{\mathit{pid}}
\newcommand{\attrs}{\mathit{attrs}}
\newcommand{\res}{\mathit{res}}
 
\usepackage{titlesec}
\titleformat{\section}[runin]
  {\normalfont\bfseries}{\thesection.}{0.5em}{}[\,---\hspace{0.35em}]
\titlespacing*{\section}{0pt}{1.1\baselineskip}{0.5em}
\titleformat{\subsection}[runin]
  {\normalfont\bfseries\itshape}{\thesubsection.}{0.4em}{}[.\hspace{0.35em}]
\titlespacing*{\subsection}{0pt}{0.7\baselineskip}{0.4em}
 
\hypersetup{
  pdftitle={Cryptographically verifiable authorization for autonomous AI agents: a falsifiable hypothesis and proof of concept},
  pdfauthor={M. Llambi-Morillas, D. Fernandez-Fernandez},
  pdfkeywords={access control, agentic security, autonomous AI agents,
    cryptographic authorization, cryptographic protocols, verifiable
    authorization, zero-knowledge proofs, zk-SNARKs}
}
 
\renewenvironment{abstract}
  {\vspace{0.4em}\begin{center}\begin{minipage}{0.90\textwidth}\small
   \setlength{\parindent}{1em}}
  {\end{minipage}\end{center}\vspace{0.3em}}

\makeatletter
\def\bstctlcite{\@ifnextchar[{\@bstctlcite}{\@bstctlcite[@auxout]}}
\def\@bstctlcite[#1]#2{\@bsphack
  \@for\@citeb:=#2\do{%
    \edef\@citeb{\expandafter\@firstofone\@citeb}%
    \if@filesw\immediate\write\csname #1\endcsname{\string\citation{\@citeb}}\fi}%
  \@esphack}
\makeatother

\begin{document}
\bstctlcite{IEEEexample:BSTcontrol} 
 
\twocolumn[%
  \begin{center}
    {\fontsize{17}{22}\selectfont\bfseries
     Cryptographically verifiable authorization for autonomous AI agents:
     a falsifiable hypothesis and proof of concept \par}
    \vspace{1.6em}

    {\large Mar~Llamb\'i-Morillas\,\orcidlink{0009-0007-2295-6256}\textsuperscript{\,1,\dag} \quad
     Daniel~Fern\'andez-Fern\'andez\,\orcidlink{0009-0005-8510-5892}\textsuperscript{\,2}\par}
    \vspace{0.5em}

    {\small\itshape
     \textsuperscript{1}Universidad Tecnol\'ogica Atl\'antico-Mediterr\'aneo (UTAMED), M\'alaga, Spain\par
     \vspace{0.2em}
     \textsuperscript{2}Universidad de Santiago de Compostela (USC), Santiago de Compostela, Spain\par}
    \vspace{0.8em}

    {\small \today\par}
    \vspace{1em}

    \begin{minipage}{0.90\textwidth}\small\itshape\centering
      This is the author version (v3) of an article published in \emph{Frontiers in Computer Science} on 24~September~2026. The Version of Record is available online at \doilink{10.3389/fcomp.2026.1966725}. Readers should consult and cite the published article: M.~Llamb\'i-Morillas and D.~Fern\'andez-Fern\'andez, ``Cryptographically verifiable authorization for autonomous AI agents: a falsifiable hypothesis and proof of concept,'' \emph{Frontiers in Computer Science}, vol.~8, art.~1966725, 2026, \doilink{10.3389/fcomp.2026.1966725}.
    \end{minipage}
  \end{center}
  \vspace{1.0em}
 
\begin{abstract}
\noindent\textbf{Abstract.} Autonomous AI agents increasingly execute actions, invoke tools, and operate on protected resources with limited human oversight. Existing authentication and authorization mechanisms establish identity and delegate authority but do not inherently provide cryptographic evidence that a concrete request issued by a specific agent satisfies the applicable policy in a specific execution context. This study hypothesizes that agent authorization can be formalized as a cryptographically verifiable relation, denoted $R_{CVA}$, that jointly binds an agent principal, a concrete authorization request, an execution context, and the satisfaction of an applicable policy, while selectively preserving the confidentiality of private authorization attributes. We introduce a preliminary formal abstraction for Cryptographically Verifiable Agent Authorization (CVA), define a compact set of candidate security properties including authorization soundness, principal binding, request binding, policy binding, and replay resistance, and provide an executable zero-knowledge proof of concept that instantiates selected elements of the model over a Groth16 zk-SNARK construction. We further identify and formalize the structural separation among identity binding, authorization-request binding, and runtime execution binding as a central open problem in the design of secure agentic systems, a distinction to our knowledge, has not been formalized within a cryptographically verifiable authorization relation by current agentic security frameworks, and present a falsifiable research agenda for its resolution.

\vspace{1.2em}
\end{abstract}
\vspace{1.4em}
]

\section{Introduction}
\label{sec:intro}
Autonomous AI agents are no longer passive software clients. They are increasingly designed to invoke tools, call APIs, access protected resources, construct multi-step action sequences, delegate subtasks, and act on external resources with limited human supervision. This transition from deterministic client behavior to autonomous, context-dependent request generation changes the security object over which authorization mechanisms must reason. In such systems, it is insufficient to establish only \emph{who} or \emph{what} an agent is, or whether a credential has been delegated; the system must determine whether a concrete request, produced by a concrete agent under a concrete execution context, satisfies an applicable policy.

Let $A_i \in A$ denotes an autonomous agent, $q_i$ an authorization request issued by $A_i$, $\alpha$ a requested action to execute, $\res$ a concrete resource to access, $c$ an execution context, and $P_j$ an applicable policy (the symbol $P$ is reserved for the set of policies, Subsection~\ref{subsec:sysreq}). The distinction motivating this study can be expressed as
\begin{equation}
  \mathrm{AuthN}(A_i) \nRightarrow \mathrm{AuthZ}(A_i,\alpha,\res,c,P_j)
  \label{eq:authn}
\end{equation}
or in its simplified form:
\begin{equation}
  \mathrm{AuthN}(A_i) \nRightarrow \mathrm{AuthZ}(q_i).
  \label{eq:authnsimple}
\end{equation}
The symbol $\nRightarrow$ \eqcite{in \mbox{Equations~\ref{eq:authn}} and~\mbox{\ref{eq:authnsimple}}} is used intentionally. It does not denote message flow, but logical non-implication: Authentication or delegation may be necessary for authorization, but neither is sufficient to establish that a concrete request is authorized.

Similarly, if a principal user $U$ delegates a capability $\kappa$ to agent $A_i$:
\begin{equation}
  \mathrm{Delegate}(U,A_i,\kappa) \nRightarrow \mathrm{AuthZ}(q_i).
  \label{eq:delegate}
\end{equation}
\eqcite{As \mbox{Equation~\ref{eq:delegate}} indicates, delegation} may establish authority over a broad class of operations, while the permissibility of a concrete request $q_i$ may still depend on the target resource ($\res \in R$), the execution context ($c$, which may encode risk level, environment classification, or session state), the applicable policy version ($\ppid_j$), and temporal validity ($t$). These attributes are defined formally as components of the authorization request in Subsection~\ref{subsec:sysreq}; they are enumerated here to make explicit that delegation alone does not determine the permissibility of any specific request.

\newpage

This observation is not isolated to the present work. Recent institutional analyses have characterized what ISACA~\cite{gupta2025iam} terms ``the looming authorization crisis'' (the systematic failure of existing IAM frameworks, including OAuth~2.0~\cite{hardt2012}, OpenID Connect, and SAML) when applied to agentic systems that violate their foundational assumptions of deterministic behavior and a single authenticated principal. The OpenID Foundation~\cite{south2025} has similarly identified the insufficiency of adapting existing protocols and proposed new identity management primitives for agentic contexts. The present work complements these analyses by providing a formal characterization of the gap: not merely as an architectural insufficiency, but as a missing security property: namely, the absence of a cryptographically verifiable relation binding a specific agent request to policy satisfaction.

This gap has become consequential only recently, as AI agents have transitioned from controlled pipelines to open agentic architectures, that is, systems in which an agent may autonomously select tools, construct multi-step action plans, and act upon external resources in ways not anticipated at deployment time. In this setting, the permissibility of a concrete request cannot be determined at authentication time but must be evaluated at request time against a policy that may depend on dynamic runtime attributes.

Recent work has advanced decentralized agent identity~\cite{liu2025diap}, accountable execution~\cite{lin2025agentid}, zero-trust agent architectures~\cite{huang2026zerotrust}, ZKP-based policy compliance~\cite{adapala2025aegis}, delegation, and privacy-preserving audit. A zero-knowledge proof (ZKP) is a cryptographic protocol in which a prover demonstrates the truth of a statement to a verifier without revealing any information beyond the validity of that statement itself~\cite{goldwasser2019}. The contribution of this study is therefore not the generic application of zero-knowledge proofs to agentic systems. Instead, we investigate a narrower and more precise question: \textbf{whether authorization itself can be represented as a request-bound cryptographically verifiable relation, distinct from identity authentication, capability delegation, and post-hoc audit}. The relevant statement is therefore not merely that an agent owns a credential, corresponds to a registered identity, or executes approved code. The statement of interest is whether a specific authorization request is supported by private attributes and contextual information that satisfy an applicable policy.

We formulate the following central hypothesis:

\smallskip
\noindent\textbf{H1---Cryptographically Verifiable Agent Authorization
Hypothesis.} \emph{Authorization decisions for autonomous AI agents can be
represented as cryptographically verifiable relations that jointly bind an agent
principal, a concrete authorization request, an execution context, and
satisfaction of an applicable policy, while selectively preserving the
confidentiality of private authorization attributes.}
\smallskip

We denote by $R_{CVA}$ the relation underlying cryptographically verifiable authorization (CVA). At a high level, $x$ encodes the public authorization statement (comprising the agent's public identifier), a commitment to the requested action, a context commitment, a policy identifier, and freshness parameters, while $w$ encodes the private witness, including the agent's secret, private authorization attributes, and the preimages of the public commitments. We define cryptographically verifiable authorization as the ability of a prover to produce evidence $\pi$ showing that a public authorization statement $x$ is supported by private witness data $w$ that satisfy an authorization relation $R_{CVA}$, without revealing $w$.

Formally, we postulate the existence of a relation:
\begin{equation}
  R_{CVA}(x,w) \in \{0,1\}.
  \label{eq:cva}
\end{equation}

Let $\lambda$ denotes the cryptographic security parameter. The proof-system parameters are generated with respect to the authorization relation:
\begin{equation}
    pp \leftarrow \mathrm{Setup}(1^{\lambda}, R_{CVA}),
    \label{eq:setup}
\end{equation}
where $pp$ abstracts the public parameter material required for proving and verification, such as circuit-specific proving and verification keys in a SNARK implementation. Let $\pi$ denotes the generated proof. Then:
\begin{equation}
  \pi \leftarrow \mathrm{Prove}(pp,x,w),
  \label{eq:prove}
\end{equation}
and a verifier accepts if:
\begin{equation}
  \mathrm{VerifyAuth}(pp,x,\pi) = 1.
  \label{eq:verify}
\end{equation}
$\mathrm{VerifyAuth}$ denotes authorization-level proof verification. Intuitively, $x$ captures what the verifier is allowed to know, while $w$ captures what the prover must know but should not disclose. VerifyAuth invokes the verifier of the underlying ZKP system over the public statement $x$ and proof $\pi$. Stateful checks, such as replay protection, are modeled separately. This proof system enables a verifier to establish that an authorization relation holds without learning the private witness, yielding the following conceptual separation:
\begin{equation}
\begin{split}
  \mathit{Identity\ Evidence} \;\not\equiv\;{}& \mathit{Authorization\ Evidence}\\
  \not\equiv\;{}& \mathit{Execution\ Evidence}.
\end{split}
\label{eq:evidence}
\end{equation}

We use the term \emph{evidence object} to denote the cryptographic, procedural, or runtime artifact presented to justify a security-relevant claim about an agent. In agentic systems, \eqcite{as \mbox{Equation~\ref{eq:evidence}} indicates,} different evidence objects support different claims. Identity evidence supports claims about who or what an agent is. Delegation evidence supports claims about authority transferred from a principal to an agent. Authorization evidence supports claims about whether a concrete request satisfies a policy under a given context. Execution evidence supports claims about what action was actually executed at runtime. Audit evidence supports claims about historical traceability after execution.

We refer to this missing object as \emph{authorization evidence}: evidence that a concrete agent request, rather than merely an agent identity or delegated credential, satisfies an applicable policy under a given context.

This study makes three contributions. First, it introduces a preliminary formal abstraction of cryptographically verifiable agent authorization as a request-bound security relation. Second, it defines a compact set of candidate properties addressing soundness, principal binding, request binding, policy binding, and replay resistance. Third, it maps an existing zero-knowledge authorization prototype to the abstraction and uses it as evidence of constructive feasibility.

\section{Related work and positioning}
\label{sec:related}
Zero-knowledge proofs have progressively moved from foundational interactive protocols~\cite{goldwasser2019} to practical non-interactive constructions suitable for real-world deployment. Groth16~\cite{groth2016} established the basis for efficient pairing-based non-interactive arguments; STARKs~\cite{bensasson2018} removed trusted-setup requirements at the cost of larger proofs; Bulletproofs~\cite{bunz2018} provided efficient range proofs without trusted setup. Systematic surveys~\cite{sheybani2025,gupta2025zkp} confirm the increasing maturity of ZKP frameworks and tooling, while applied work demonstrates their viability in distributed authentication contexts~\cite{naitcherif2023,mo2025,xiang2025,gopal2025}.

In the agentic domain, DIAP~\cite{liu2025diap} addresses decentralized agent identity and ownership via ZKP over a hybrid P2P stack. Huang et al.~\cite{huang2026zerotrust} investigate zero-trust identity architectures and fine-grained access control for agentic AI. Binding Agent ID~\cite{lin2025agentid} advances user-to-agent-to-code binding with execution provenance. The Aegis Protocol~\cite{adapala2025aegis} incorporates ZKP-based policy compliance into a broader agent security architecture combining decentralized identifiers and post-quantum cryptography. Adjacent work explores privacy-preserving audit of agent communications~\cite{guanlin2025} and verifiable delegation in agent interaction protocols. These works establish the cryptographic feasibility of proving private statements without disclosure; however, the authorization object studied here is not a generic statement but a request-specific relation involving principal, request, context, and policy. 

These systems demonstrate that agent identity, execution provenance, delegation, and policy compliance can be made verifiable. The remaining question addressed in this study is whether authorization itself can be isolated as an independent cryptographic relation.

The resulting landscape motivates an explicit separation among identity, delegation, policy compliance, audit, ZKP-based verification, and request-specific authorization, as summarized in Table~\ref{tab:positioning}.

\begin{table*}[t]
\centering
\caption{Positioning of related work with respect to authorization-relevant
security properties.}
\label{tab:positioning}
\small
\begin{tabularx}{\textwidth}{@{}lYYYYYY@{}}
\toprule
\textbf{Work} & \textbf{Identity} & \textbf{Delegation} &
\textbf{Policy compliance} & \textbf{Audit} &
\textbf{Pre-execution authorization} & \textbf{ZKP}\\
\midrule
DIAP~\cite{liu2025diap}        & \yes              & \nomark    & \nomark       & \nomark     & \nomark  & \yes\\
Huang et al.~\cite{huang2026zerotrust}& \yes              & Partial    & \yes          & \nomark     & Partial  & Partial\\
BAID~\cite{lin2025agentid}       & \yes              & \nomark    & Partial       & \nomark     & Partial  & \yes\\
Aegis~\cite{adapala2025aegis}      & \yes              & \nomark    & \yes          & \nomark     & Partial  & \yes\\
zk-MCP~\cite{guanlin2025}     & \nomark           & \nomark    & Rule-oriented & \yes        & \nomark  & \yes\\
AIP~\cite{prakash2026}        & \yes              & \yes       & Attenuation   & Provenance  & Partial  & \notmark\\
\textbf{Proposed CVA}   & Principal binding, not full identity framework & Orthogonal & \yes & \nomark & \textbf{\yes} & \textbf{\yes}\\
\bottomrule
\end{tabularx}
\end{table*}

The intended gap is deliberately narrow: existing study substantially advances identity, delegation, policy compliance, and audit, whereas \textbf{request-bound cryptographically verifiable authorization remains insufficiently characterized as an independent security abstraction}. This study focuses specifically on the relation among principal, request, context, and policy. Table~\ref{tab:positioning} reveals that while ZKP-based approaches have been applied to agent identity, code binding, policy compliance, and audit, no existing work explicitly models authorization as a request-bound cryptographically verifiable relation. This motivates the abstraction introduced in Section~\ref{sec:model}. As noted in Section~\ref{sec:intro}, this claim concerns the explicit formalization of authorization as a request-bound cryptographic relation, not the novelty of applying zero-knowledge proofs to agentic systems in general, which DIAP, Aegis, and Huang et al. already establish (see Table~\ref{tab:positioning}).

Adjacent work by Prakash~\cite{prakash2026} introduces Invocation-Bound Capability Tokens for verifiable delegation across MCP and A2A, demonstrating sub-millisecond verification overhead in reference implementations evaluated in a real multi-agent deployment. While AIP advances the delegation and identity layers with cryptographic token chains, it does not model authorization as a request-bound relation with verifiable policy satisfaction and private witness; the two contributions address complementary layers of the agentic trust stack.

The general phenomenon underlying this gap is not new: It is structurally related to the classical confused-deputy problem~\cite{confuseddeputyhardy1988} and to time-of-check-to-time-of-use (TOCTOU) race conditions~\cite{bishop1996checking} and has recently been shown to manifest concretely in LLM-enabled agents~\cite{lilienthal2025mind}. To our knowledge, however, no existing work formalizes this gap (the question of whether the action authorized at verification time is the same action subsequently executed) as a structural property of a cryptographically verifiable authorization relation, distinguishing authorization-request binding from runtime execution binding within a ZKP-based construction. This structural observation, formalized in Subsection~\ref{subsec:execbind}, is not explicitly addressed in current agentic security frameworks and motivates a distinct line of formal inquiry.

\section{Preliminary formal model}
\label{sec:model}

\subsection{System and request model}
\label{subsec:sysreq}
We model an agentic authorization environment as a tuple:
\begin{equation}
  S = (A,G,T,R,P),
  \label{eq:system}
\end{equation}
where $A$ is the set of autonomous agents, $G$ the set of authorization gateways or verifiers, $T$ the set of external tools or callable services, $R$ the set of protected resources, and $P$ the set of authorization policies. To refer to a specific protected resource \eqcite{in \mbox{Equation~\ref{eq:system}}}, we write $\res \in R$. Unless otherwise stated, the model developed in this section addresses a single agent-gateway authorization interaction; its extension to multi-hop delegation chains is discussed as a structural limitation in Section~\ref{sec:discussion}.

Let $\lambda$ denotes the cryptographic security parameter. The notation $\{0,1\}^{\lambda}$ denotes the set of bitstrings of length $\lambda$. An agent $A_i \in A$ holds private key material denoted as $\msk$:
\begin{equation}
  \msk_i \in \{0,1\}^{\lambda},
  \label{eq:sk}
\end{equation}
and exposes a public identity commitment:
\begin{equation}
  \aid_i = \mathrm{CommitID}(\msk_i,\rho_i)
  \label{eq:idcommit}
\end{equation}
where $\mathrm{CommitID}$ denotes a randomized cryptographic commitment scheme for agent identity, $\msk_i$ \eqcite{in \mbox{Equation~\ref{eq:sk}}} is the private secret of $A_i$, and $\rho_i$ is the commitment randomness, with $\rho_i \in \{0,1\}^{\lambda}$. In standard cryptographic notation, this corresponds to a commitment of the form $\mathrm{Com}(m;r)$. This formulation~\cite{pedersen1992} is consistent with privacy-preserving and anonymous authentication protocols, where commitments allow a prover to bind to secret information without disclosing it. \eqcite{The commitment-based identifier of \mbox{Equation~\ref{eq:idcommit}}} allows the model to represent agent binding without requiring direct disclosure of the underlying secret material. In the proof-of-concept implementation, this abstraction is instantiated in simplified form as $\aid_i = \mathrm{Poseidon}(\msk_i)$~\cite{grassi2021}, primarily for ZK-circuit efficiency. This provides principal binding under hash assumptions but does not realize the randomized hiding semantics of the general commitment-based model;
this distinction is acknowledged as a limitation.

We define an authorization request as
\begin{equation}
  q_i = (\aid_i,\alpha,\res,c,\ppid_j,n,t)
  \label{eq:request}
\end{equation}
where the tuple $q_i$ is designed to be minimal but sufficient: $\aid_i$ commits to the requesting agent $A_i$, $\alpha$ is the requested action, $\res \in R$ identifies the concrete protected resource, $c$, the execution context, captures runtime-dependent attributes that may affect policy evaluation, $n$ is a nonce, and $t$ is a validity epoch or timestamp. We reserve $P$ for the set of authorization policies. A concrete policy is denoted as $P_j \in P$, following the same indexing convention as $A_i \in A$. The versioned policy identifier $\ppid_j = H_P(\mathit{policy\_code}_j \parallel \mathit{version}_j)$ (not to be confused with the proof $\pi$ introduced in Equation~\ref{eq:prove}) is a compact cryptographic reference to $P_j$, used in the public statement to bind the proof to a specific policy version without encoding the full policy in the circuit. Additional attributes such as risk scores or session identifiers may be incorporated into $c$ without modifying the core model structure.

An action may be decomposed as
\begin{equation}
  \alpha = (\mathit{tool}, \mathit{operation}, \mathit{args}), \quad \mathit{tool} \in T.
  \label{eq:action}
\end{equation}
Where request data are sensitive, the public commitment is computed as
\begin{equation}
  h_q = H_q(\mathrm{Encode}(\mathit{domain}, \mathit{tool}, \mathit{operation}, \res, \mathit{args})),
\label{eq:hq}
\end{equation}
where $\mathrm{Encode}$ denotes a deterministic canonical serialization function.
We denote by
\begin{equation}
   q_{priv} := (\mathit{domain},\mathit{tool},\mathit{operation},\res,\mathit{args}) 
   \label{eq:qpriv}
\end{equation}
the private canonical representation of the request committed in Equation~\ref{eq:hq}; the execution context $c$ is committed separately through $h_c$ (see Subsection~\ref{subsec:stmtwit}), while the remaining fields of $q_i$ ($\aid_i$, $\ppid_j$, $n$, $t$) appear in the clear in the public statement. Canonical encoding \eqcite{of the representation in \mbox{Equation~\ref{eq:qpriv}}} is required to avoid ambiguity between semantically equivalent but byte-distinct requests. Without canonical encoding, two semantically equivalent requests may yield different commitments, while two differently serialized requests may create ambiguity about what was actually authorized. Throughout the study, $H_q$, $H_c$, and $H_P$ denote domain-separated collision-resistant hash functions, or domain-separated invocations of the same underlying hash primitive. The subscript identifies the semantic domain of the commitment: request, context, or policy. The lowercase values $h_q$, $h_c$, and $\ppid_j$ denote the resulting public commitments or identifiers.

\subsection{Public statement and private witness}
\label{subsec:stmtwit}
The public statement is defined as
\begin{equation}
  x = (\aid_i,h_q,h_c,\ppid_j,n,t).
  \label{eq:x}
\end{equation}
As established in Subsection~\ref{subsec:sysreq}, $h_q = H_q(\mathrm{Encode}(q_{priv}))$ is the public commitment to the authorization request, as defined in Equation~\ref{eq:hq}. Similarly, $h_c = H_c(\mathrm{Encode}(c_{priv}))$ is a commitment to security-relevant contextual attributes, where $H_c$ is a collision-resistant hash function applied to a canonical serialization of the private context. The inclusion of $h_c$ in the public statement extends the binding surface of the relation beyond the request itself to encompass runtime-dependent conditions (such as environment classification, resource state, or session attributes) that may affect policy evaluation. While $h_c$ is part of the formal model, its implementation is left open in the current proof of concept and identified as a priority for future implementation.

The private witness is
\begin{equation}
  w = (\msk_i,\rho_i,\attrs_i,q_{priv},c_{priv})
  \label{eq:w}
\end{equation}
where $\attrs_i$ encodes private authorization attributes not revealed to the verifier, $q_{priv}$ denotes the private canonical request known by the prover, while $h_q$ is its public commitment. Similarly, $c_{priv}$ denotes private contextual attributes, while $h_c$ is their public commitment. The value $\rho_i$ \eqcite{in \mbox{Equation~\ref{eq:w}}} is the randomness required to open the identity commitment $\aid_i$.

The separation between $x$ and $w$ captures the privacy objective of the model: the verifier observes the statement required for authorization, while sensitive attributes and preimages remain private.

\subsection{Core authorization relation}
\label{subsec:relation}
The preliminary CVA relation is defined as $R_{CVA}(x,w) = 1$ if and only if the
following constraints are jointly satisfied:
\begin{align}
  \mathrm{CommitID}(\msk_i,\rho_i) &= \aid_i, \label{eq:c15}\\
  H_q(\mathrm{Encode}(q_{priv})) &= h_q \label{eq:c16}
\end{align} 
and
\begin{align}
 H_c(\mathrm{Encode}(c_{priv})) &= h_c \label{eq:c17}
\end{align}
where $h_c$ is the context commitment introduced in Equation~\ref{eq:x} and
\begin{equation}
  P_j(\attrs_i,q_{priv},c_{priv}) = 1.
  \label{eq:c18}
\end{equation}
Here each policy $P_j \in P$ is modeled as a deterministic predicate over $(\attrs_i, q_{priv}, c_{priv})$, and $P_j$ is the policy referenced by the identifier $\ppid_j$ contained in the public statement $x$.
In compact form:
\begin{equation}
\begin{split}
  R_{CVA} = \;&\mathrm{BindPrincipal} \land \mathrm{BindRequest}\\
  &\land\, \mathrm{BindContext} \land \mathrm{SatisfyPolicy}.
\end{split}
\label{eq:rcva}
\end{equation}
We define BindPrincipal as
\begin{equation}
  \mathrm{BindPrincipal} := [\,\mathrm{CommitID}(\msk_i,\rho_i) = \aid_i\,].
  \label{eq:bindprincipal}
\end{equation}
The intended proof of knowledge is
\begin{equation}
  \pi = \mathrm{PoK}\{\,(\msk_i,\rho_i,\attrs_i,q_{priv},c_{priv}):
  R_{CVA}(x,w) = 1\,\}.
\label{eq:pok}
\end{equation}
Equivalently, \eqcite{\mbox{Equation~\ref{eq:pok}} states that} $\pi$ proves knowledge of a witness $w$ satisfying $R_{CVA}(x,w) = 1$. Each conjunct in $R_{CVA}$ rules out a distinct class of transfer attack. Omitting $\mathrm{BindPrincipal}$\eqcite{, \mbox{Equation~\ref{eq:bindprincipal}},} enables cross-principal reuse; omitting $\mathrm{BindRequest}$ enables policy-satisfying proofs to authorize unintended actions; omitting $\mathrm{BindContext}$ weakens context-dependent authorization; and omitting $\mathrm{SatisfyPolicy}$ decouples the proof from the authorization decision itself. Their conjunction is the minimal condition for request-specific authorization verifiability.

This construction proves satisfaction of a committed-request relation; it does not prove the internal reasoning or deliberative process of the agent:
\begin{equation}
  \mathit{Request\ Commitment} \neq \mathit{Internal\ Agent\ Intent}.
  \label{eq:intent}
\end{equation}
\eqcite{\mbox{Equation~\ref{eq:intent}} therefore delimits the semantic scope of the evidence produced.}
Freshness depends on external mutable state and is therefore separated from the stateless cryptographic core:
\begin{equation}
  \mathrm{Accept}(x,\pi) = \mathrm{VerifyAuth}(pp,x,\pi) \land \mathrm{Fresh}(n,t,N)
  \label{eq:accept}
\end{equation}
where $N$ is the set of previously consumed nonces maintained by the gateway.

\subsection{Threat model}
\label{subsec:threat}
We consider a probabilistic polynomial-time adversary $\mathrm{Adv}$ operating against the CVA authorization workflow. The adversary may observe, intercept, and replay authorization statements and proofs; substitute or forge public agent identifiers; modify or substitute request commitments; present proofs under policy contexts different from those for which they were generated; and control unauthorized or partially authorized agents. The adversary cannot break the assumed cryptographic primitives, directly obtain honest party secrets, or compromise the trusted authorization gateway. Although gateway compromise falls outside the adversary's assumed capabilities, we note for robustness that, where the gateway compromised, freshness enforcement (see Subsection~\ref{subsec:replay}) would be voided, since nonce-set and timestamp-window state are maintained gateway-side. However, principal binding, authorization-request binding, and policy binding remain governed by the cryptographic core of Equations~\ref{eq:c15}--\ref{eq:c18} and are not weakened by gateway compromise, since these guarantees depend on SNARK verification over the public statement, not on gateway-held state.

These capabilities identify five primary attack classes against the proposed abstraction. First, proof forgery: the adversary attempts to produce an accepting proof for a statement for which no valid witness exists, addressed by Authorization Soundness (see Subsection~\ref{subsec:soundness}). Second, cross-principal transfer: the adversary attempts to reuse a valid proof generated for principal $\aid_i$ to authorize a request under a different principal $\aid_k$, addressed by Principal Binding (see Subsection~\ref{subsec:principal}). Third, cross-request transfer: the adversary attempts to reuse a proof bound to request $q_i$ to authorize a distinct request $q_i'$, addressed by Authorization-Request Binding (see Subsection~\ref{subsec:request}). Fourth, cross-policy transfer: the adversary attempts to present a proof generated under one policy identifier as evidence under another policy identifier, addressed by Policy Binding in Equation~\ref{eq:policybind}. Fifth, replay: the adversary reuses a previously accepted proof to obtain a second authorization, addressed by Replay Resistance (see Subsection~\ref{subsec:replay}). The goal of this adversary model is not to cover all failures of autonomous execution but to isolate attacks that target authorization evidence itself. A complete symbolic adversary model for multi-agent authorization chains is left as future work.

\section{Candidate security properties}
\label{sec:properties}
The following properties are intended to address the principal attack vectors identified in the threat model of Subsection~\ref{subsec:threat}: proof forgery without a valid witness, cross-principal proof transfer, cross-request proof transfer, cross-policy proof transfer, and proof reuse. Complete reductions of these properties to standard cryptographic assumptions, in particular the soundness and knowledge-soundness of the underlying proof system and the binding property of $\mathrm{CommitID}$, are left as part of the future research agenda. The separation between authorization-request binding and runtime execution binding in Subsection~\ref{subsec:execbind} is included as a structural observation rather than a game-based property since it identifies a limitation of the ZKP authorization layer itself. Two further properties assessed in Table~\ref{tab:coverage} are not stated as separate games: Attribute privacy is inherited directly from the zero-knowledge property of the underlying proof system, and context binding is captured structurally as the conjunct (Equation~\ref{eq:c17}) of $R_{CVA}$, with its transfer-resistance form given in Equation~\ref{eq:contextbind}. Accordingly, each property formalized in Subsections~\ref{subsec:soundness}--\ref{subsec:replay} is stated as a candidate definition: Its status as an established security guarantee, rather than a well-motivated design target, remains contingent on the reduction arguments discussed at the end of this section.

\subsection{Authorization soundness}
\label{subsec:soundness}
Let $\mathrm{Adv}$ denotes a probabilistic polynomial-time adversary; the adversary should not be able to produce an accepted proof for a statement for which no valid witness exists. The adversary wins the authorization soundness game $\mathrm{Game}^{\text{auth-sound}}_{\mathrm{Adv}}(\lambda)$ if it outputs $(x^{\star},\pi^{\star})$ such that:
\begin{equation}
  \mathrm{VerifyAuth}(pp,x^{\star},\pi^{\star}) = 1 \land \nexists\, w : R_{CVA}(x^{\star},w) = 1.
\label{eq:soundgame}
\end{equation}
A secure construction requires
\begin{equation}
  \mathrm{Adv}^{\text{auth-sound}}_{\mathrm{Adv}}(\lambda) \leq \mathrm{negl}(\lambda).
  \label{eq:soundbound}
\end{equation}
\eqcite{The bound of \mbox{Equation~\ref{eq:soundbound}}} follows from two assumptions. First, the underlying proof system must satisfy knowledge-soundness: An accepting proof for a statement should imply the existence of an extractable witness for the encoded relation. Second, the implemented circuit and verification metadata must correctly encode and bind the intended relation $R_{CVA}$. Thus, an adversary that wins the game in Equation~\ref{eq:soundgame} must either violate knowledge-soundness of the proof system or exploit a mismatch between the intended CVA relation and its encoded implementation. A complete reduction, including explicit extractor and simulation arguments, is left for future work.

\subsection{Principal binding}
\label{subsec:principal}
We use $\Pr[\,\cdot\,]$ to denote the probability of an event over the randomness of the adversary, the proof system, and the experiment. Let
$x_i = (\aid_i,h_q,h_c,\ppid_j,n,t)$ and $x_k = (\aid_k,h_q,h_c,\ppid_j,n,t)$ with $\aid_i \neq \aid_k$. A proof $\pi_i$ generated for principal $\aid_i$ must not verify under $x_k$:
\begin{equation}
  \Pr[\,\mathrm{VerifyAuth}(pp,x_k,\pi_i) = 1\,] \leq \mathrm{negl}(\lambda).
  \label{eq:principal}
\end{equation}
Principal binding relies on the binding property of $\mathrm{CommitID}$ and on the soundness of the proof system.

\subsection{Authorization-request binding}
\label{subsec:request}
Let $q_i \neq q_i'$ be two distinct requests whose private canonical
components differ, with commitments $h_q = H_q(\mathrm{Encode}(q_{priv}))$ and $h_q' = H_q(\mathrm{Encode}(q_{priv}'))$. Let $\pi_q$ denotes a proof generated for the statement $x_q$ containing $h_q$, and $x_{q'}$ the statement containing $h_q'$. A proof bound to $q_i$ must not authorize $q_i'$:
\begin{equation}
  \Pr[\,\mathrm{VerifyAuth}(pp,x_{q'},\pi_q) = 1\,] \leq \mathrm{negl}(\lambda).
  \label{eq:reqbind}
\end{equation}
This property is central to the proposed abstraction: abstract policy satisfaction alone is insufficient if the authorization evidence can be detached from the specific request to which it applies.

A related condition, which we term Policy Binding, establishes that a proof
generated under policy version $\ppid_j$ must not verify as valid evidence under a
distinct policy version $\ppid_k$:
\begin{equation}
\begin{split}
  &\ppid_j \neq \ppid_k \Rightarrow{} \\
  &\Pr[\,\mathrm{VerifyAuth} (pp,x_{\ppid_k},\pi_{\ppid_j}) = 1\,] \leq \mathrm{negl}(\lambda).
\label{eq:policybind}
\end{split}
\end{equation}
This property is enforced by including $\ppid_j$ in the public statement $x$ and constraining the circuit to evaluate the policy identified by $\ppid_j$ as defined in Subsection~\ref{subsec:sysreq}.

An analogous condition, which we term \textbf{Context Binding}, requires that a proof generated under one context commitment must not verify under a statement carrying a distinct context commitment. Let
\begin{equation}
    h_c = H_c(\operatorname{Encode}(c_{\mathrm{priv}}))
    \label{eq:hc}
\end{equation}
and
\begin{equation}
    h'_c = H_c(\operatorname{Encode}(c'_{\mathrm{priv}})), 
    \qquad c_{\mathrm{priv}} \neq c'_{\mathrm{priv}}.
    \label{eq:hzetac}
\end{equation}
Under collision resistance of $H_c$, $c_{\mathrm{priv}} \neq 
c'_{\mathrm{priv}}$ implies $h_c \neq h'_c$\eqcite{, as given in \mbox{Equations~\ref{eq:hc}} and~\mbox{\ref{eq:hzetac}}}. Let
\begin{equation}
    x_c = (id_i,\, h_q,\, h_c,\, pid_j,\, n,\, t)
    \label{eq:xc}
\end{equation}
and
\begin{equation}
    x_{c'} = (id_i,\, h_q,\, h'_c,\, pid_j,\, n,\, t).
    \label{eq:xczeta}
\end{equation}
A proof $\pi_c$ generated for $x_c$ \eqcite{in \mbox{Equation~\ref{eq:xc}}} must not verify under $x_{c'}$ \eqcite{in \mbox{Equation~\ref{eq:xczeta}}}:
\begin{equation}
    \Pr\bigl[\mathsf{VerifyAuth}(pp,\, x_{c'},\, \pi_c) = 1\bigr] 
    \leq \mathsf{negl}(\lambda).
    \label{eq:contextbind}
\end{equation}
This property relies on the collision resistance of $H_c$, canonical context encoding, and the correct inclusion of $h_c$ as a constrained  public input of $R_{\mathrm{CVA}}$. Context Binding prevents substitution of the context commitment at authorization time; it does not, by itself, prevent context drift between authorization and  subsequent execution, which is addressed separately as a \textsc{TOCTOU} limitation in Section~\ref{sec:discussion}.

\subsection{Replay resistance}
\label{subsec:replay}
Let $N$ denotes the consumed nonce set defined in Equation~\ref{eq:accept}. Once a request with nonce $n$ has been successfully accepted, the gateway updates its replay-control state as follows:
\begin{equation}
  N \leftarrow N \cup \{n\},
  \label{eq:nonceupdate}
\end{equation}
The complete freshness predicate is defined as
\begin{equation}
   \mathrm{Fresh}(n,t,N) = 1 \Leftrightarrow (n \notin N) \land (t_{\min} \leq t \leq t_{\max})
  \label{eq:fresh1}
\end{equation}
\begin{equation}
  \mathrm{Fresh}(n,t,N) = 0 \Leftrightarrow (n \in N)\ \lor\ (t < t_{\min}) \lor\ (t > t_{\max})
  \label{eq:fresh0}
\end{equation}
\begin{equation}
  n \in N \Rightarrow \mathrm{Fresh}(n,t,N) = 0.
  \label{eq:freshreuse}
\end{equation}

Equation~\ref{eq:fresh1} defines acceptance: a nonce is fresh if it has not been previously consumed and the timestamp falls within the authorized validity window, delimited by the gateway-defined bounds $t_{\min}$ and $t_{\max}$. Equation~\ref{eq:fresh0} captures both replay via nonce reuse and replay via deferred presentation outside the validity window as rejection conditions. Subsequent reuse of $n$ is therefore rejected, as in Equation~\ref{eq:freshreuse}. Equations~\ref{eq:fresh0}, \ref{eq:freshreuse} are the explicit contrapositive forms of Equation~\ref{eq:fresh1}, stated separately to make the rejection conditions directly enforceable at the gateway.

Replay resistance is a property of the complete authorization workflow and depends on stateful external enforcement at the gateway. The consequence of $\mathrm{Fresh}(n,t,N) = 0$ is that the full acceptance predicate of Equation~\ref{eq:accept} evaluates to false, regardless of proof validity. The ZK proof binds the request to the public statement, while the gateway enforces single-use semantics through mutable nonce state \eqcite{of \mbox{Equation~\ref{eq:nonceupdate}}} and temporal validity checks. Note that gateway-side freshness is meaningful only because $n$ and $t$ are bound to the proof as components of the public statement $x$: A proof not bound to its nonce could be re-presented by an adversary under a fresh nonce $n' \notin N$, voiding replay control. Statement binding of $n$ and $t$ is therefore a precondition of Subsection~\ref{subsec:replay}, not an optional design choice.

\subsection{Authorization binding is not equivalent to execution binding}
\label{subsec:execbind}
A valid proof establishes that a committed request satisfies the authorization relation:
\begin{equation}
  \mathrm{VerifyAuth}(pp,x_q,\pi) = 1.
  \label{eq:pocverify}
\end{equation}
This does not imply that the runtime subsequently executes the same request. Formally,
\begin{equation}
  \mathrm{VerifyAuth}(pp,x_q,\pi) = 1 \nRightarrow q_{exec} = q_{auth}
  \label{eq:execnot}
\end{equation}
where $q_{auth}$ denotes the request committed in the verified statement $x_q$, and $q_{exec}$ denotes the request actually executed by the runtime. Therefore,
\begin{equation}
\begin{split}
  \mathit{Authorization\ Request\ Binding}\\
  \not\equiv \ \mathit{Runtime\ Execution\ Binding}.
\end{split}
\label{eq:authneqexec}
\end{equation}
\eqcite{The gap recorded in \mbox{Equation~\ref{eq:authneqexec}}} is structural rather than incidental. A ZK circuit is evaluated at proof generation time over a committed representation of the intended request; it has no access to the runtime state at execution time and cannot constrain what the agent subsequently does with the authorization it receives. Closing this gap requires a trust anchor that operates at execution time rather than at proof generation time. Establishing runtime consistency requires an additional trusted evidence mechanism, such as remote attestation, a trusted execution environment, or verifiable execution receipts, beyond the guarantees of the ZKP authorization relation. This constitutes an open problem and is identified as part of the research agenda.

\subsection{Informal reduction sketches}
\label{subsec:informalreduction}
The following arguments sketch, at an informal level, how a successful adversary against each of the four properties above could be used to construct an adversary against a standard cryptographic assumption. Since Groth16 is knowledge-sound~\cite{groth2016}, it is in particular sound: The existence of a knowledge extractor for accepting proofs directly implies that no accepting proof exists for a false statement except with negligible probability. We write $\mathrm{Adv}^{\text{sound}}_{\text{Groth16}}(\lambda)$ to denote the advantage of a probabilistic polynomial-time algorithm in producing an accepting Groth16 proof for a false statement, $\mathrm{Adv}^{\text{coll}}_{H}(\lambda)$ to denote the advantage of finding a collision under a hash function $H$, and $\mathrm{Adv}^{\text{bind}}_{\mathsf{CommitID}}(\lambda)$ to denote the advantage of finding an alternate opening of $\mathsf{CommitID}$. We invoke knowledge-soundness directly in Subsection~\ref{subsec:soundness}, where witness extraction is required, and rely on the weaker, directly implied soundness property in the remaining sketches, where only the non-existence of an accepting proof for an altered public input is needed. Consistent with Limitation 1 in Section~\ref{sec:limitations}, these sketches are informal and do not include an explicit extractor or simulator construction; a complete formal treatment remains future work.

\begin{itemize}[leftmargin=1.1em,itemsep=0.55em,topsep=0.45em]

\item \textbf{Authorization Soundness} (Subsection~\ref{subsec:soundness}). Let $\mathrm{Adv}$ be the adversary defined in Subsection~\ref{subsec:soundness}, winning the game of Equation~\ref{eq:soundgame} with non-negligible advantage. By knowledge-soundness of Groth16, there exists a probabilistic polynomial-time extractor $\mathsf{Ext}$ that, given the same input distribution as $\mathrm{Adv}$,
\begin{equation}
\begin{split}
\Pr\big[\, R_{CVA}(x^\star, w) = 1 \;:\;
 &w \leftarrow \mathsf{Ext}(x^\star, \pi^\star) \,\big]\\
 &\geq\; 1 - \mathrm{negl}(\lambda).
\end{split}
\label{eq:extractor}
\end{equation}
If extraction succeeds \eqcite{as in \mbox{Equation~\ref{eq:extractor}}}, then $R_{CVA}(x^\star, w) = 1$ by correctness of the circuit encoding, contradicting the assumption that no valid witness exists for $x^\star$; if extraction fails, $\mathrm{Adv}$ directly yields an adversary against knowledge-soundness of Groth16. Either branch bounds $\mathrm{Adv}$'s advantage by the knowledge-soundness advantage of Groth16, up to the caveat, already noted at the end of Subsection~\ref{subsec:soundness}, that the circuit must correctly encode the intended relation $R_{CVA}$.

\item \textbf{Principal Binding} (Subsection~\ref{subsec:principal}). An adversary winning the game of Equation~\ref{eq:principal} can succeed via one of two avenues. First, it may present a proof $\pi_i$, honestly generated for the statement $x_i$ containing $\aid_i$, as evidence under a statement $x_k$ differing only in the identity coordinate ($\aid_k \neq \aid_i$). Since Groth16's verification equation is a fixed linear combination over the public inputs, accepting the same proof under two distinct values of a public-input coordinate directly contradicts soundness of Groth16, provided $\aid_i$ is correctly wired as a distinguishing public input in the circuit (the same encoding-correctness caveat as above). Second, the adversary may attempt to find an alternate opening $(\msk_i', \rho_i') \neq (\msk_i, \rho_i)$ with $\mathsf{CommitID}(\msk_i', \rho_i') = \aid_i$, i.e., break the binding property of the identity commitment; since $\mathsf{CommitID}$ follows the Pedersen construction~\cite{pedersen1992}, this reduces directly to the hardness of the discrete logarithm problem in the underlying group. Both avenues combine into the following bound:
\begin{equation}
\begin{split}
\mathrm{Adv}_{\mathrm{Adv}}^{\text{principal-bind}}(\lambda) \;\leq\;
 &\mathrm{Adv}^{\text{sound}}_{\text{Groth16}}(\lambda)\\
 &+\; \mathrm{Adv}^{\text{bind}}_{\mathsf{CommitID}}(\lambda).
\end{split}
\label{eq:principal-bind-reduction}
\end{equation}

\item \textbf{Authorization-Request Binding and Policy Binding} (Subsection~\ref{subsec:request}). By the same public-input rigidity argument \eqcite{that produced \mbox{Equation~\ref{eq:principal-bind-reduction}}}, consider an adversary winning either the request-binding game of Equation~\ref{eq:reqbind} or the policy-binding game of Equation~\ref{eq:policybind}. In either case, the adversary breaks soundness of Groth16 with respect to the corresponding public-input coordinate ($h_q$ or $\ppid_j$, respectively) or breaks collision-resistance of the corresponding hash function ($H_q$ or $H_P$). Both properties therefore satisfy the same bound, up to substitution of the coordinate and hash function \eqcite{in \mbox{Equation~\ref{eq:request-policy-bind-reduction}}}:
\begin{equation}
\begin{split}
\mathrm{Adv}_{\mathrm{Adv}}^{\text{req/policy-bind}}(\lambda) \;\leq\;
 &\mathrm{Adv}^{\text{sound}}_{\text{Groth16}}(\lambda) \;+\; \mathrm{Adv}^{\text{coll}}_{H}(\lambda),\\
 &H \in \{H_q, H_P\}.
\end{split}
\label{eq:request-policy-bind-reduction}
\end{equation}

\item \textbf{Replay Resistance} (Subsection~\ref{subsec:replay}). This property is structurally different from the three above and should not be read as reducing to a single hardness assumption. Its cryptographic component follows the same public-input rigidity argument applied to the $(n, t)$ coordinates of the public statement: detaching a proof from its bound nonce and timestamp, and re-presenting it under fresh values, breaks soundness of Groth16 with respect to those coordinates. Its remaining component is not cryptographic but procedural: Given that $(n, t)$ are correctly bound to the proof, replay resistance additionally requires that the gateway correctly and continuously enforces the freshness predicate of Equations~\ref{eq:fresh1}--\ref{eq:freshreuse} over its nonce set $N$. As already noted in Subsection~\ref{subsec:threat}, this second component depends on gateway integrity and falls outside the cryptographic core of the construction:
\begin{equation}
\begin{split}
\mathrm{Adv}_{\mathrm{Adv}}^{\text{replay}}(\lambda) \;\leq\;
 &\mathrm{Adv}^{\text{sound}}_{\text{Groth16}}(\lambda)\\
 &+\; \varepsilon_{\text{gateway}},
\end{split}
\label{eq:replay-reduction}
\end{equation}
where $\varepsilon_{\text{gateway}}$ denotes the probability that the gateway fails to correctly maintain or enforce the freshness state. If the gateway is compromised, this term \eqcite{of \mbox{Equation~\ref{eq:replay-reduction}}} --but not principal, request, or policy binding-- is voided.

\end{itemize}

\section{Proof-of-concept implementation}
\label{sec:poc}

\subsection{Scope}
\label{subsec:scope}
To assess the constructive feasibility of selected elements of the CVA abstraction, an executable zero-knowledge authorization prototype was developed and publicly presented at RootedCON Madrid~\cite{llambi2026a}, an independent security conference, prior to the formalization introduced in this study~\cite{llambi2026b}. This artifact is the experimental precursor from which the present hypothesis emerged.

\begin{figure*}[h!]
\centering
\includegraphics[width=\textwidth]{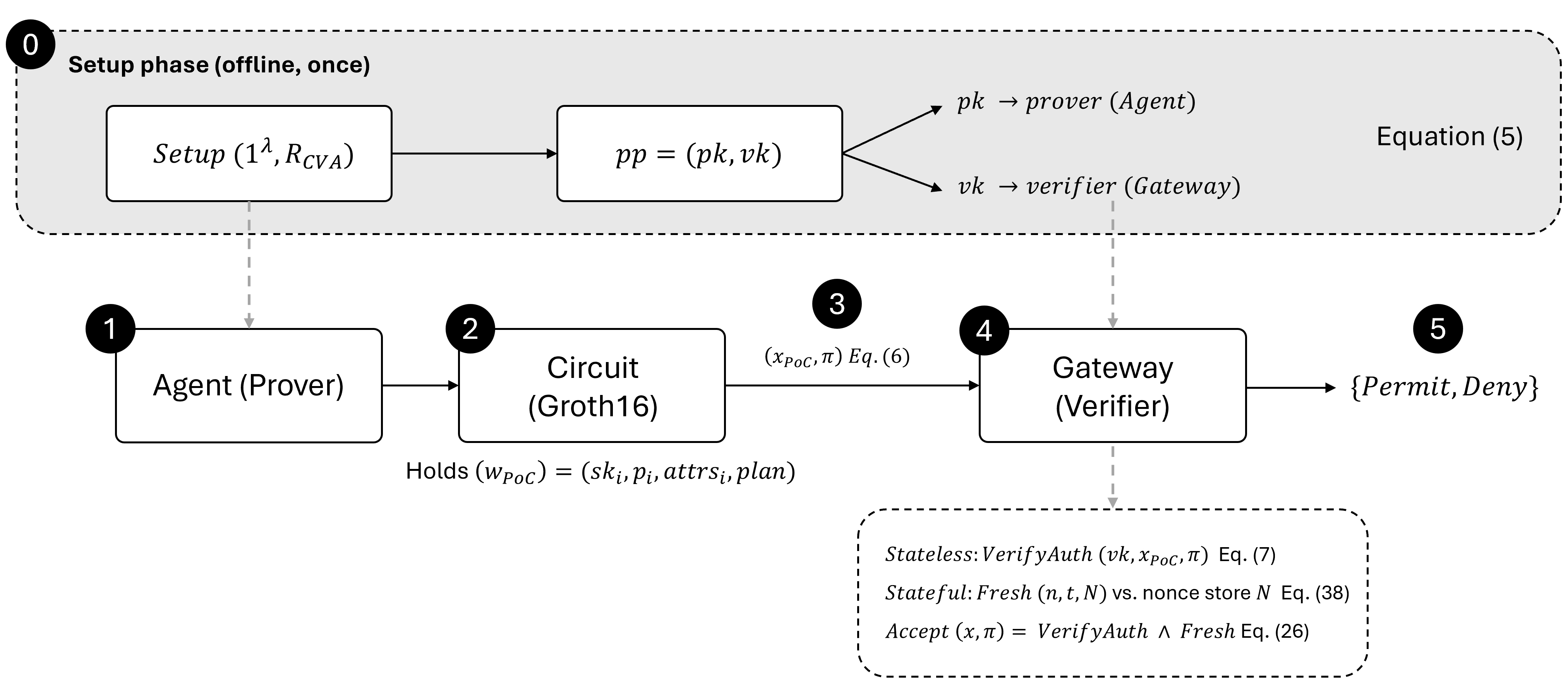}
\caption{Proof-of-concept authorization workflow, showing the offline setup phase and the runtime authorization flow. (0) A one-time setup ceremony produces the proving and verification key pair $pp = (pk,vk)$ via Equation~\ref{eq:setup}; $pk$ is provided to the prover (Agent) and $vk$ to the verifier (Gateway). (1) The Agent, acting as prover, holds the private witness (see Equation~\ref{eq:wpoc}). (2) The Circuit (Groth16) evaluates the arithmetic constraints described in Subsection~\ref{subsec:protorel} over this witness. (3) The resulting public statement and proof $(x_{PoC},\pi)$ are transmitted to the Gateway (see Equation~\ref{eq:prove}). (4) The Gateway, acting as verifier, evaluates a stateless proof-verification check (see Equation~\ref{eq:verify}) together with a stateful freshness check against its nonce store $N$ (see Equation~\ref{eq:fresh1}); both must hold for acceptance (see Equation~\ref{eq:accept}). (5) The Gateway issues an authorization decision, $\{Permit,Deny\}$. Circuit execution occurs locally, on the prover's side; the verifier never accesses the circuit itself, only the verification key $vk$.}
\label{fig:flow}
\end{figure*}

The prototype is not presented as a complete validation of the CVA abstraction. It demonstrates that selected bindings and private policy predicates can be instantiated in an executable pre-authorization workflow, providing constructive evidence of feasibility for the principal- and plan-level binding conjuncts of the CVA abstraction. The architecture follows the conceptual flow shown in Figure~\ref{fig:flow}.

The implementation employs the Groth16 zk-SNARK construction~\cite{groth2016} over the bn128 curve, with circuits written in Circom 2.x and proof generation via snarkjs. Groth16 was selected for its minimal proof size and constant verification time, properties favorable for gateway-side enforcement. The trusted setup requirement and the absence of post-quantum security are acknowledged limitations; alternative schemes are identified as part of the comparative evaluation agenda in Section~\ref{sec:discussion}. The gateway is implemented as a FastAPI service performing stateless proof verification and stateful replay control.

\subsection{Prototype relation}
\label{subsec:protorel}
The proof of concept instantiates a restricted version of the general CVA relation introduced in Section~\ref{sec:model}. As noted in Subsection~\ref{subsec:sysreq}, the formal model uses a randomized commitment $\mathrm{CommitID}(\msk_i,\rho_i)$; the prototype instantiates this using a randomized two-input Poseidon hash rather than a Pedersen-style commitment, for ZK-circuit efficiency. Therefore, the prototype should be interpreted as an implementation of principal binding and request binding under a hash-based hiding assumption, not as a full realization of the discrete-logarithm-based generalized commitment model.

The public statement verified by the gateway is defined as
\begin{equation}
  x_{PoC} = (\aid_i,h_{plan},n,t)
  \label{eq:xpoc}
\end{equation}
where $\aid_i$ denotes the public identifier of the agent, $h_{plan}$ is the public commitment to the plan or requested action sequence, $n$ is a nonce, and $t$ is the validity timestamp or epoch. \eqcite{The public statement of \mbox{Equation~\ref{eq:xpoc}}} contains only the information required for the verifier to bind the proof to a specific agent, a specific committed plan, and a freshness window; in particular, $n$ and $t$ are bound to the proof as public inputs of the verified statement, as required by the replay precondition of Subsection~\ref{subsec:replay}.

The corresponding private witness is
\begin{equation}
  w_{PoC} = (\msk_i,\rho_i,\attrs_i,\mathit{plan})
  \label{eq:wpoc}
\end{equation}
where $\msk_i$ and $\rho_i$ are the private secret and commitment randomness of the agent jointly opening the identity commitment of Equation~\ref{eq:poseidon}, $\attrs_i$ denotes private authorization attributes, and $\mathit{plan}$ represents the action sequence or request structure whose public commitment is $h_{plan}$. These values remain private to the prover and are not disclosed to the gateway.

The prototype relation is defined as
\begin{equation}
  R_{PoC}(x_{PoC},w_{PoC}) = 1
  \label{eq:rpoc}
\end{equation}
if the following three constraints on \eqcite{\mbox{Equation~\ref{eq:rpoc}}} are satisfied.

First, the prover must demonstrate knowledge of the private secret corresponding to the public agent identifier. Instantiating Equation~\ref{eq:c15},
\begin{equation}
  \mathrm{Poseidon}(\msk_i, \rho_i) = \aid_i
  \label{eq:poseidon}
\end{equation}
this constraint instantiates principal binding in the prototype using the randomized two-input construction $\mathsf{Poseidon}(\msk_i, \rho_i)$. Unlike the deterministic single-input form used in an earlier iteration of the prototype, this construction provides randomized hiding under a hash-based assumption on Poseidon (informally, that Poseidon behaves as a random oracle with respect to $\rho_i$), distinct from the discrete-logarithm-based hiding of the general Pedersen-style $\mathsf{CommitID}$ construction discussed in Subsection~\ref{subsec:sysreq} and invoked in the reduction sketch of Subsection~\ref{subsec:informalreduction}. The two constructions satisfy the same functional binding requirement but rest on different cryptographic assumptions; we do not claim the discrete-logarithm reduction of Subsection~\ref{subsec:informalreduction} applies to this concrete instantiation.

The randomized two-input construction adds 102 constraints (531 vs. 429) relative to the earlier single-input form, an increase attributable primarily to the additional Poseidon input together with an unrelated policy-check correction applied in the same revision; this overhead is negligible relative to total circuit size and does not measurably affect the witness- or proof-generation times reported in Table~\ref{tab:benchmarks} (see Subsection~\ref{subsec:performance}).

Second, the proof must be bound to the plan that the gateway is expected to authorize. As a partial implementation of Equation~\ref{eq:c16},
\begin{equation}
  \mathrm{SHA256}(\mathit{plan}) = h_{plan}
  \label{eq:sha}
\end{equation}
this constraint\eqcite{, \mbox{Equation~\ref{eq:sha}},} provides request binding at the level of the committed plan. It prevents a proof generated for one plan from being reused for a different committed plan, assuming collision resistance of the hash function and canonical encoding of the plan.

Third, the private authorization attributes and the committed plan must satisfy the policy encoded in the circuit. Instantiating Equation~\ref{eq:c18}
\begin{equation}
  P_{PoC}(\attrs_i,\mathit{plan}) = 1.
  \label{eq:ppoc}
\end{equation}
This condition\eqcite{, \mbox{Equation~\ref{eq:ppoc}},} represents the policy-satisfaction component of the prototype. The predicate $P_{PoC}$ is implemented as arithmetic constraints in the ZK circuit and therefore captures only the policy logic explicitly encoded at circuit generation time.

Freshness and replay protection are enforced outside the circuit by the authorization gateway:
\begin{equation}
  \mathrm{Fresh}(n,t,N) = 1
  \label{eq:freshpoc}
\end{equation}
where $N$ is the gateway-maintained set of previously consumed nonces. \eqcite{The separation of \mbox{Equation~\ref{eq:freshpoc}}} is intentional: Proof verification is stateless, whereas replay resistance requires mutable enforcement state. Consequently, freshness is treated as a property of the complete authorization workflow rather than as a guarantee provided solely by the ZK circuit.

In compact form, the proof of concept instantiates
\begin{equation}
\begin{split}
  R_{PoC} = & \mathrm{BindPrincipal}_{\mathrm{Poseidon}} \land
    \\ & \mathrm{BindPlan}_{\mathrm{SHA256}} \land \mathrm{SatisfyPolicy}_{\mathrm{Circuit}}
\label{eq:rpoccompact}
\end{split}
\end{equation}
with replay resistance enforced externally by the gateway as defined in Equations~\ref{eq:fresh1}, \ref{eq:fresh0}.

The PoC \eqcite{of \mbox{Equation~\ref{eq:rpoccompact}}} therefore supports the constructive plausibility of principal binding, plan-level request binding, and private policy satisfaction, while leaving full context binding and runtime execution binding outside the implemented scope. The prototype should therefore be read as an executable witness of the abstraction's feasibility, not as an empirical demonstration that the abstraction is complete.

\subsection{Property coverage assessment}
\label{subsec:coverage}
\begin{table*}[h!]
\centering
\caption{Mapping between formal candidate properties and prototype mechanisms.}
\label{tab:coverage}
\small
\begin{tabularx}{\textwidth}{@{}l X X@{}}
\toprule
\textbf{Property} & \textbf{Prototype mechanism} & \textbf{Status}\\
\midrule
Authorization soundness   & Circuit constraints and SNARK soundness assumptions & Partial\\
Principal binding         & Agent secret and randomness bound via a two-input Poseidon hash & Implemented\\
Request binding           & Plan/action commitment via SHA-256 & Implemented for plan-level binding; partial relative to full CVA request binding\\
Policy binding            & Fixed circuit/policy relation; no explicit $\ppid_j$ public input in the current prototype & Partial / not fully implemented\\
Replay resistance         & Nonce and external gateway state & Implemented externally\\
Attribute privacy         & ZK witness under Groth16 assumptions & Implemented under proof-system assumptions\\
Context binding           & No explicit context commitment & Open\\
Runtime execution binding & No trusted execution evidence & Open\\
\bottomrule
\end{tabularx}
\end{table*}

The conservative characterization is intentional. The purpose of Table~\ref{tab:coverage} is to expose the distance between the preliminary formal model and the current artifact rather than to overstate prototype coverage.

\subsection{Preliminary validation strategy}
\label{subsec:validation}
Property-oriented validation derives test cases directly from the candidate properties. Representative cases include the following: (i) a valid principal issuing a policy-compliant request; (ii) an invalid agent secret; (iii) substitution of the public identity commitment; (iv) modification of the request hash; (v) private attributes failing the encoded policy predicate; (vi) reuse of a previously consumed nonce; and (vii) tampering with public signals. These cases provide structured coverage of the instantiated properties but do not constitute complete empirical security evaluation against arbitrary adversaries.

\subsection{Preliminary performance characterization}
\label{subsec:performance}
To ground the operational-viability discussion of Section~\ref{sec:discussion} (RQ3) in concrete data rather than qualitative expectation alone, Table~\ref{tab:benchmarks} reports circuit size and timing characteristics of the current prototype instantiation.

\begin{table}[t]
\centering
\caption{Proof-of-concept performance characterization (mean $\pm$ standard deviation, $n=20$ runs, single circuit configuration: 531 constraints, 537 wires, nine private inputs, and five outputs). All timings have been measured via command-line invocation and therefore include Node.js process startup overhead; they are not library-level microbenchmarks and should not be compared directly against figures obtained via in-process API calls. Measured on an AMD Ryzen 7 7700 (8-core, 3.8\,GHz base) with 32\,GB RAM, using the WASM prover backend.}
\label{tab:benchmarks}
\small
\begin{tabularx}{\columnwidth}{@{}l X@{}}
\toprule
\textbf{Metric} & \textbf{Value} \\
\midrule
Circuit constraints & 531 \\
Witness generation time & $61.9 \pm 1.1$ ms \\
Proof generation time & $351.9 \pm 26.6$ ms \\
Verification time & $309.4 \pm 22.6$ ms \\
Proof size & 805 bytes \\
\bottomrule
\end{tabularx}
\end{table}

These figures characterize a single circuit configuration and are not intended to demonstrate empirical scaling behavior across constraint counts; the theoretical constant-time verification property of Groth16 with respect to circuit size is an established property of the proof system, not a claim this benchmark sets out to validate independently. Instructions and the underlying measurement script are included in the associated repository~\cite{llambi2026b}.

\section{Discussion and research agenda}
\label{sec:discussion}
The purpose of the preliminary model is not only to propose a construction but also to identify where the guarantees provided by ZKP-based authorization end.

The preliminary model supports four principal observations.

\textbf{First}, the separation among binding layers:
\begin{equation}
\begin{split}
  \mathit{Identity\ Binding}& \not\equiv \mathit{Authorization\ Request\ Binding} \\ \not\equiv &\mathit{Runtime\ Execution\ Binding}.
\label{eq:bindinglayers}
\end{split}
\end{equation}
A system may correctly authenticate an agent yet fail to bind authorization evidence to a specific request. It may likewise bind evidence to a request without guaranteeing that the same request is executed at runtime. These are structurally distinct security properties requiring separate mechanisms.

\textbf{Second}, the correctness of a proof must not be confused with the normative correctness of the encoded policy. If the implemented relation $R_{CVA}$ does not faithfully encode the intended organizational policy $P_{intended}$ (that is, if $R_{CVA}(x,w) = 1$ as defined in Equation~\ref{eq:cva} does not imply $P_{intended}(\attrs_i,q_{priv},c_{priv}) = 1$), then a valid proof may correspond to an incorrectly authorized decision. Policy governance, versioning via $\ppid_j$, circuit governance, and verification-key distribution therefore constitute first-class security concerns orthogonal to the cryptographic guarantees of the proof system.

\textbf{Third}, dynamic execution introduces a time-of-check-to-time-of-use problem. If verification occurs at time $t_v$ and execution at $t_e > t_v$, the relevant context may change:
\begin{equation}
  c_{t_v} \neq c_{t_e}.
  \label{eq:context}
\end{equation}
\eqcite{Given the context drift of \mbox{Equation~\ref{eq:context}}, let} $\mathrm{Authorized}(q,c)$ denotes the predicate that holds when request $q$ is authorized under context $c$; formally, when $\mathrm{Accept}(x_q,\pi) = 1$ for a proof $\pi$ generated under context $c$, as defined in Equation~\ref{eq:accept}. Consequently,
\begin{equation}
  \mathrm{Authorized}(q,c_{t_v}) = 1 \nRightarrow \mathrm{Authorized}(q,c_{t_e}) = 1.
  \label{eq:toctou}
\end{equation}
\eqcite{\mbox{Equation~\ref{eq:toctou}} is thus a context-specific instance of the general authorization-execution gap already identified in \mbox{Equation~\ref{eq:authneqexec}}: whereas \mbox{Equation~\ref{eq:execnot}} concerns drift in the identity of the executed request, (\mbox{Equation~\ref{eq:toctou}}) concerns drift in the context under which a fixed request was authorized.}

\textbf{Fourth}, a structural observation concerns multi-agent delegation chains. The present model defines the CVA relation for a single agent-gateway interaction, where a prover $A_i$ generates a proof for a single authorization request $q_i$. In practice, autonomous agents increasingly operate in orchestrated architectures where $A_i$ may delegate a subtask to another agent $A_k$, which may in turn delegate further. In such chains, the authorization question becomes compositional: $A_k$ must not only prove that its local request $q_k$ satisfies a CVA relation, but also that the delegation from $A_i$ to $A_k$ preserves, constrains, or attenuates the authorization scope granted upstream.

The current conjuncts of $R_{CVA}$ in Equation~\ref{eq:rcva}, including $\mathrm{BindPrincipal}$, $\mathrm{BindRequest}$, and $\mathrm{SatisfyPolicy}$, remain necessary but are not sufficient for multi-hop authorization soundness. A proof that $A_k$ satisfies a local policy does not by itself establish that $A_k$ was authorized to act under the scope delegated by $A_i$, or that scope has not been expanded across hops. Addressing this gap would require an additional delegation-binding component, such as $\mathrm{BindDelegationScope}$, together with either (i) recursive proof composition---in which each hop proves that it extends and constrains the previous authorization state---or (ii) a delegation-aware witness structure encoding the relevant authorization chain. Informally, extending Equation~\ref{eq:rcva} to a delegation chain $A_1 \rightarrow A_2 \rightarrow \dots \rightarrow A_n$ would require each hop $k$ to jointly satisfy its local relation and a delegation-binding condition $\text{scope}_k \subseteq \text{scope}_{k-1}$, ensuring that authorization scope is preserved or narrowed, never expanded, across hops. We deliberately leave this condition unformalized here as its full formalization falls outside the scope of the present work and is reserved for the compositional security analysis identified as future work.

These four observations directly motivate the research questions that structure the future empirical and formal agenda:

\smallskip
\noindent\textbf{RQ1---Necessary Bindings.} Which dimensions must be cryptographically bound to prevent proof transfer across principals, requests, contexts, policies, and execution domains? In particular, the present model includes context binding through $h_c$ as a formal conjunct in $R_{CVA}$, but its implementation and the characterization of which contextual attributes are necessary and sufficient for policy-relevant binding remain open. RQ1 subsumes this as a central sub-question. A candidate resolution to the implementation gap noted in RQ1 involves one of two constructions. First, fixed-length canonical padding with domain separation: context attributes are serialized into a fixed-size vector prior to hashing, with unused slots padded to a canonical null value, avoiding the ambiguity that variable-length inputs would otherwise introduce into the arithmetic circuit. Second, a Poseidon-sponge-based construction supporting variable-length absorption, following the general sponge paradigm, would allow $h_c$ to commit to a context of arbitrary length without a fixed upper bound at circuit-compilation time, at the cost of additional constraint overhead per absorbed block. Both constructions remain unimplemented in the current prototype and are proposed here as design directions rather than validated mechanisms.

\smallskip
\noindent\textbf{RQ2---Representable Policies.} Which classes of agent authorization policies can be efficiently represented as arithmetic ZK relations under practical circuit complexity constraints?

\smallskip
\noindent\textbf{RQ3---Operational Viability.} What overhead is introduced by request-bound ZKP authorization under different proof systems and agent workload profiles?

\smallskip
Preliminary evidence from the proof of concept suggests that, for the selected Groth16 implementation \cite{groth2016}, gateway-side verification remains effectively constant with respect to circuit constraint count (see Table~\ref{tab:benchmarks} for measured verification time on the current prototype instantiation). In contrast, proof generation scales with the number of arithmetic constraints and is therefore expected to dominate end-to-end authorization latency as policy expressiveness increases. Operational viability consequently depends not only on the selected proof system but also on the authorization frequency of the deployment model. A low-frequency model, in which a proof is generated once per task plan and verified once at the gateway, presents a fundamentally different latency budget from a high-frequency model in which each tool invocation triggers an independent authorization cycle. RQ3 should therefore be evaluated across at least three axes: proof-system selection, circuit complexity as a function of policy expressiveness, and authorization frequency as a function of the agent workload profile. This distinction also reinforces the separation between authorization evidence and runtime execution evidence: Reducing authorization frequency may improve latency, but it increases the importance of binding the authorized plan to subsequent execution.

These observations suggest that CVA should be evaluated along two separate axes: cryptographic validity of the authorization evidence, and operational consistency between authorization and execution. The first can be studied through proof-system soundness, binding properties, and circuit correctness; the second requires additional runtime evidence mechanisms and cannot be obtained from the authorization proof alone.

\section{Limitations}
\label{sec:limitations}
We explicitly acknowledge the following limitations:
\begin{enumerate}
  \item The formal model is preliminary; complete security reductions for the proposed candidate properties to standard cryptographic assumptions remain as future work. \vspace{-0.1cm}
  \item The prototype has not been subjected to independent cryptographic circuit audit. \vspace{-0.1cm}
  \item Groth16 requires a trusted setup ceremony; the resulting common reference string constitutes an external trust assumption. The system is additionally not post-quantum secure under current assumptions. \vspace{-0.1cm}
  \item The encoded policy language is constrained to static arithmetic circuits; dynamic or procedural policies are not representable without circuit recompilation. \vspace{-0.1cm}
  \item Context binding is included in the formal model through $H_c(\mathrm{Encode}(c_{priv})) = h_c$ but is not implemented in the current prototype. \vspace{-0.1cm}
  \item Runtime execution binding is an open problem not addressed by the ZKP authorization layer alone, as formalized in Equations~\ref{eq:pocverify}, \ref{eq:execnot}. \vspace{-0.1cm}
  \item No comparative empirical benchmark across proof systems has been conducted. \vspace{-0.1cm}
  \item No multi-agent or multi-principal delegation chain has been evaluated. \vspace{-0.1cm}
  \item The authorization gateway is partially trusted; its compromise is outside the threat model of the core cryptographic construction. \vspace{-0.1cm}
  \item The prototype's request binding uses a single plan-level commitment (SHA256(plan)) rather than the model's decomposition of the request into distinct resource (see Equation~\ref{eq:request}) and action components (see Equation~\ref{eq:action}). \vspace{-0.1cm}
\end{enumerate}

\section{Conclusion}
\label{sec:conclusion}
This study argues that authorization for autonomous AI agents should be studied as a cryptographically verifiable security relation, rather than solely as a consequence of authenticated identity or delegated credentials. We hypothesize that a meaningful class of authorization decisions can be represented through relations jointly binding an agent principal, a concrete authorization request, an execution context, and the satisfaction of an applicable policy, while selectively preserving private authorization attributes.

Table~\ref{tab:positioning}, discussed in Section~\ref{sec:related}, positions the proposed abstraction relative to existing work: while prior contributions advance identity, delegation, policy compliance, and audit, the CVA abstraction targets the orthogonal property of request-bound verifiable authorization, which remains uncharacterized as an independent security relation in the surveyed literature.

To support this hypothesis, we introduced a preliminary formal model, defined a compact set of candidate security properties, and mapped an executable zero-knowledge proof of concept to selected elements of the abstraction. The artifact is presented as constructive evidence of feasibility for the implemented principal- and request-binding conjuncts, rather than complete formal or empirical validation of the full CVA relation.

The central conclusion is the explicit separation defined in Equation~\ref{eq:bindinglayers}:
\begin{equation*}
\begin{split}
  \mathit{Identity\ Binding}& \not\equiv \mathit{Authorization\ Request\ Binding} \\ \not\equiv &\mathit{Runtime\ Execution\ Binding}.
\end{split}
\end{equation*}

This distinction clarifies both the contribution and the limits of zero-knowledge authorization: a proof may establish that a committed request satisfies an encoded policy but cannot by itself establish that the same request was subsequently executed. The resulting abstraction provides a falsifiable basis for future formal security analysis, comparative evaluation across proof systems, and integration with trusted runtime execution evidence.

Beyond the specific research questions identified in Section~\ref{sec:discussion}, the CVA abstraction suggests a broader reorientation of how authorization is studied in agentic systems: not as a configuration problem solved at deployment time but as a per-request cryptographic property that must be established, verified, and composed across agent interactions. As autonomous agents become infrastructure (invoking tools, delegating tasks, and acting upon protected resources at scale), the gap between authenticated identity and verified authorization is likely to widen. Closing it requires not only formal foundations of the kind proposed here but also a broader effort to develop authorization evidence as a first-class security primitive alongside identity, delegation, and runtime execution evidence.

\vspace{0.5cm}
\hrule
\small

{\footnotesize
\section*{Data availability statement}
Publicly available datasets were analyzed in this study. This data can be found here: \url{https://github.com/Imari91/zk-auth-agent-demo}.
\vspace{-0.1cm}

\section*{Author contributions}
ML-M: Investigation, Software, Conceptualization, Writing -- review \& editing, Formal analysis, Writing -- original draft, Methodology. DF-F: Writing -- review \& editing, Validation, Supervision.
\vspace{-0.1cm}

\section*{Funding}
The author(s) declared that financial support was not received for this work and/or its publication.
\vspace{-0.1cm}

\section*{Conflict of interest}
The author(s) declared that this work was conducted in the absence of any commercial or financial relationships that could be construed as a potential conflict of interest.
\vspace{-0.1cm}

\section*{Generative AI statement}
The author(s) declared that Generative AI was used in the creation of this manuscript. Claude Opus 5 (Anthropic, \url{https://claude.ai}) was used on specific passages for English-language editing and proofreading. The manuscript was written entirely by the author(s), who reviewed and edited all output and take full responsibility for the content of the publication.
}

\newpage

\nocite{
  gupta2025iam,
  hardt2012,
  south2025,
  liu2025diap,
  lin2025agentid,
  huang2026zerotrust,
  adapala2025aegis,
  goldwasser2019,
  groth2016,
  bensasson2018,
  bunz2018,
  sheybani2025,
  gupta2025zkp,
  naitcherif2023,
  mo2025,
  xiang2025,
  gopal2025,
  guanlin2025,
  prakash2026,
  confuseddeputyhardy1988,
  bishop1996checking,
  lilienthal2025mind,
  pedersen1992,
  grassi2021,
  llambi2026a,
  llambi2026b}

{\fontsize{9}{10}\selectfont
\bibliographystyle{IEEEtran}
\bibliography{references}
}

\blfootnote{\textsuperscript{\dag}\,Corresponding author:
\href{mailto:mar.llambi@gmail.com}{mar.llambi@gmail.com}}

\end{document}